# Using Earth to Search for Long-Range Spin-Velocity Interactions

N. B. Clayburn*, L. R. Hunter
Physics Department, Amherst College, Amherst, Massachusetts 01002, USA

Precision measurements of the possible coupling of spin to other (pseudo)scalar and (pseudo)vector fields has proven to be a sensitive way to search for new particle physics beyond the standard model. Indeed, in addition to searching for exotic spin-spin interactions, studies have been undertaken to look for couplings of spin to gravity, the relative velocity between particles, and preferred directions. Several laboratory experiments have established upper bounds on the energy associated with various fermion spin-orientations relative to Earth. Here, we combine these results with a model of Earth as a moving unpolarized source in order to investigate the possible long-range spin-velocity interactions associated with the exchange of ultralight ($m_{z'} < 1$ neV) or massless scalar or vector bosons. We establish stringent bounds on the strength of these couplings between electrons, neutrons, protons, and nucleons.

## I. INTRODUCTION

The Standard Model of particle physics describes established forces as being mediated by the exchange of virtual bosons. Many standard model extensions predict the existence of new bosons [1]. Axions and axion-like particles (ALPS) [2, 3, 4] are examples of proposed spin-0 bosons while $z'$ bosons, paraphotons, and dark photons [5, 6, 7, 8] are examples of proposed spin-1 bosons. All of these bosons are potential dark-matter candidates. Also of interest are proposed extensions of general relativity, like torsion gravity [9], which could give rise to a coupling of spin to mass.

Here we consider the possible existence of ultralight or massless spin-0 or spin-1 bosons that mediate exotic long-range interactions between spin and velocity. The exchange of such a particle between fermions can be modeled as an effective potential. A general classification of interactions between non-relativistic fermions that assumes only rotational invariance gives 16 possible operator structures [10]. A more complete analysis which describes phenomena that arise on the (sub)atomic scale is presented in Ref. [11]. That description includes contact terms which are important for atomic-scale phenomena but can be ignored at long range. Here we restrict our discussion to the "spin-velocity" interactions:

$$V_{4+5} = -f_{4+5} \frac{\hbar^2}{8\pi m_1 c} [\hat{\sigma}_1 \cdot (\vec{v} \times \hat{r})] \left(\frac{1}{\lambda r} + \frac{1}{r^2}\right) e^{-r/\lambda}, \quad (1)$$

and

$$V_{12+13} = f_{12+13} \frac{\hbar}{8\pi} [\hat{\sigma}_1 \cdot \vec{v}] \frac{1}{r} e^{-r/\lambda}. \quad (2)$$

These potentials describe the possible interactions between a polarized particle of mass $m_1$ with spin $\vec{s}_1 = \hbar \hat{\sigma}_1/2$ interacting with a unpolarized particle. Here $\vec{r}$ and $\vec{v}$ denote the relative position and relative velocity between the two particles and $\hat{r} = \vec{r}/r$ is the corresponding unit vector. Here, $f_{4+5}$ and $f_{12+13}$ are dimensionless coupling constants, $\hbar$ is the reduced Planck's constant, $c$ is the speed of light, and $\lambda = \hbar/(m_{z'} c)$ is the interaction range where $m_{z'}$ is the intermediate-boson mass. We note that while both of these potentials are even under a time-reversal operation (T), the potential $V_{12+13}$ is odd under a parity operation (P). Both of these potential structures accommodate well-known physics. The potential $V_{4+5}$, for the exchange of an ordinary virtual photon, describes the potential ($V = -\vec{\mu} \cdot \vec{B}$) of a particle with a magnetic moment $\vec{\mu}$ interacting with the magnetic field $\vec{B}$ produced by a charge $e$ moving with velocity $\vec{v}$. The potential $V_{12+13}$ can describe the effective potential associated with the exchange of the weak neutral boson, $Z_0$. Here we are interested in exploring if these same potential forms might also describe the exchange of other exotic bosons.

*nclayburn@amherst.edu 

## A. Previous Results

In order to probe these potentials, a relative velocity between a detection spin and an unpolarized source is required. In recent investigations of these potentials using atomic magnetometers [12, 13, 14, 15] the source was a modulated laboratory mass. An earlier spin-polarized torsion-pendulum experiment [16] used both the Sun and Moon as moving sources in their analysis. Previous work in our lab combined a spin-polarized model of Earth with the results of several experiments to extract stringent long-range bounds on the possible coupling strengths of both velocity-independent and velocity- dependent spin-spin interactions [17, 18]. In Ref. [18] it is pointed out that Earth's rotation creates substantial relative velocities between typical Earth- atoms and laboratory spins. Another analysis used Earth as a unpolarized mass source to place bounds on the couplings $f_{12+13}$ [19] and a number of works have used Earth as either a polarized [20] or unpolarized source [16, 21, 22] to constrain monopole-dipole couplings. In the present work we use Earth as a velocity-dependent unpolarized source. We combine our Earth model with equations 1 and 2 and the results of various experiments to obtain bounds on the long-range spin-velocity coupling constants $f_{4+5}$ and $f_{12+13}$.

Exotic spin-velocity dependent interactions have been experimentally probed over a broad range of interaction-lengths and boson-mass scales. We use the convention that the first fermion from each labeled pair represents the spin-polarized particle and the second fermion represents the unpolarized particle. We follow the standard shorthand of representing the electron, proton, neutron, and nucleon as e, p, n, and N, respectively. Previous e-N bounds for both the $V_{4+5}$ [12, 13, 14] and $V_{12+13}$ [13, 15] potentials for $10^{-2}$ m $< \lambda < 10^7$ m were established using atomic magnetometers. At interaction lengths above $10^7$ m torsion-pendulum experiments [16] have set limits on e-N couplings for both potentials. For the same range, bounds have recently been placed on spin-velocity n-N couplings by combining the motions of the Sun and Moon with the results of an experiment that searched for a violation of Local Lorentz Invariance [23]. For $\lambda < 10^{-2}$ m, constraints on $V_{4+5}$ n-N couplings have been established using K-Rb-$^{21}$Ne co-magnetometers and a tungsten ring featuring a high nucleon density [24], a spin-based amplifier [25], and a slow neutron polarimeter [26]. In the same range, p-N constraints for $V_{4+5}$ were also established by Ref. [24]. Previous $V_{12+13}$ n-N bounds were established by analyzing the spin-relaxation time of polarized $^3$He gas [19].

At interaction-lengths below $10^{-2}$ m, constraints of $V_{12+13}$ e-N couplings are set by atomic parity non-conservation experiments [27], magnetic force microscopy [28], and single NV centers [29]. The $V_{4+5}$ e-N coupling constraints are set by stellar cooling limits [30] and cantilever experiments [31]. An analysis of these bounds is presented in Ref [32]. An unreviewed result [33] claims an observation of an interaction consistent with potential $V_{4+5}$ and the existence of two new exotic bosons with masses of about 0.6 eV and 25 meV.

## II. METHODS

Here we use the Earth as a moving unpolarized source. The geoparticles can interact (via the proposed anomalous spin-velocity potential) with electrons or nucleons contained in spin-sensitive detectors. These interactions can induce energy shifts in the detection spins that depend on their orientation with respect to Earth. The experimental signature of these potentials is the reversibility of the potential with the reversal of the detector spin orientation or the applied magnetic field. The average velocity of the majority of Earth's mass relative to a detector at Earth's surface is towards the west. Consequently, the spin-dot-velocity $V_{12+13}$ potential is maximally sensitive to orientation-dependent energy shifts with spins oriented east/west while the $V_{4+5}$ cross-product potential is more sensitive to energy shifts with spins oriented north/south. Because Earth's surface speed is greater at the equator than the poles, the latitude of the experiment also influences the experiment's sensitivity. For example, at a latitude of 42.37° (Amherst, MA) one's speed around Earth's center is ~343 m/s. For comparison, laboratory sources typically have velocities of less than ~6 m/s.

Using Earth as a particle source requires creating a map of the electron and nucleon densities everywhere within Earth. Two input functions of the distance from the center of Earth, $r'$, are needed to create this map: the mass density of Earth, $\rho_m(r')$, and the number of particles per unit mass $\kappa_j(r')$ of the considered fermion species ($j$). By multiplying these two parameters one obtains the particle density of each species at $r'$.

For the Earth as a nucleon source, the potentials of Eq. 1 and Eq. 2 can be formulated as an effective total potential, $V_T$, acting on the polarized spins [32]:

$$V_{T,4+5} = -f_{4+5} \frac{\hbar^2}{8\pi m_1 c} \int_{Earth} \rho(r')[\hat{\sigma}_1 \cdot (\vec{v} \times \hat{r})] \left(\frac{1}{\lambda r} + \frac{1}{r^2}\right) e^{-\frac{r}{\lambda}} d^3 r' \quad (3)$$

and

$$V_{T,12+13} = f_{12+13} \frac{\hbar}{8\pi} \int_{Earth} \rho(r')[\hat{\sigma}_1 \cdot \vec{v}] \frac{1}{r} e^{-\frac{r}{\lambda}} d^3 r'. \quad (4)$$

Here $\rho(r') = \kappa_N(r')\rho_m(r')$ is Earth's unpolarized nucleon particle density. We use the coordinate system detailed in Ref. [18] to describe the integration of the potentials over Earth's volume. The potentials of Eq. 3 and Eq. 4 are evaluated at $\vec{r} = \vec{r}'_A - \vec{r}'$, where the geoparticle location is described by the vector $\vec{r}'$ and the location of the spin-polarized particle is designated by the vector $\vec{r}'_A$. The relative velocity between the geoparticle and spin-polarized particle is $\vec{v} = \vec{v}'_A - \vec{v}'$ where $\vec{v}'_A = \vec{\Omega} \times \vec{r}'_A$ and $\vec{v}' = \vec{\Omega} \times \vec{r}'$. Here, $\vec{\Omega}$ is the angular rotation vector of Earth with magnitude $|\vec{\Omega}| = 2\pi/(1$ sidereal day). The integration is carried out on the Amherst College computer cluster in geocentric coordinates using Mathematica.

For $\lambda > 10^4$ m we use the preliminary reference Earth model (PREM) to determine Earth's density. This model describes Earth's average properties by depth and is widely used as the basis for seismic tomography and related global geophysical models [34]. In this long-range limit $\kappa_j$ is determined from the elemental composition of the various Earth strata. We assume an elemental composition determined by a smooth layered model which describes the composition of the core [35], mantel [36], crust, and ocean [37]. We assume natural isotopic abundance when determining the number of nucleons associated with each element. Atoms inside Earth are assumed neutral in charge such that the number of protons and electrons for each atom are equal. In total when averaging over Earth, the ratio between protons and nucleons is ~0.486.

This model sufficiently describes Earth's properties for large $\lambda$, but for small $\lambda$ local inhomogeneities must be considered. We use a global crustal model, CRUST 1.0 [38] to describe Earth's density for $\lambda < 10^4$ m. The model includes 8 layers: water, ice, 3 sediment layers and upper, middle and lower crystalline crust. At these short ranges we assume that protons and neutrons occur in equal quantities. We find good agreement between bounds for intermediate $\lambda$ values resulting from the two models and our predicted bounds are markedly insensitive to changes in various Earth-model parameters. We do not quote bounds for $\lambda < 10^2$ m as inhomogeneities near the experiment limit their accuracy.

### III. RESULTS

We examine the experiments which place the most stringent bounds on these orientation-dependent energy shifts ($\beta$) for electrons and nucleons. Bounds on the electron ($e$) energy when its spin is oriented north ($N$) and east ($E$) are derived from the SmCo$_5$-Alnico torsion-pendulum experiment [16]. The bounds on the neutron ($n$) and proton ($p$) orientation-dependent energy with spins oriented north [39] and east [17] are derived from $^{199}$Hg-$^{133}$Cs co-magnetometer experiments. To extract these bounds it is assumed that $^{199}$Hg has a neutron-spin projection of -31% and a proton spin projection of -3% [40, 41]. Other experiments intended to search for an anomalous scalar coupling between nucleon spin and Earth's gravity yield bounds for neutron [22] and proton [21] spin orientations along Earth's spin axis ($z$). These results are most effective at bounding the $V_{4+5}$ potential. The $^{129}$Xe-$^{131}$Xe co-magnetometer experiment of Ref. [22] and the $^{85}$Rb-$^{87}$Rb co-magnetometer experiment of Ref. [21] measure the energy difference between spin-up and spin-down states of their associated particles along the vertical to be $<5.3 \times 10^{-22}$ eV and $<3.4 \times 10^{-18}$ eV, respectively (95% C.L). These experiments place bounds on the neutron and proton spin coupling along the $z$-axis, $\hat{\beta}^n_z < (5.3 \times 10^{-22}$ eV$)\cos(31.82°)/2 = 2.3 \times 10^{-22}$ eV and $\beta^p_z < (3.4 \times 10^{-18}$ eV$)\cos(37.66°)/2 = 1.3 \times 10^{-18}$ eV, respectively. Here the geometrical factor transforms the bound from along the vertical axis to along the $z$-axis and the factor of 2 accounts for the two orientations of the nuclear spins with respect to the applied magnetic field.

| Ref. | System | $\beta$ (eV) | Location | Spin | Symbol |
|------|--------|--------------|----------|------|--------|
| [16] | AlNiCo-SmCo$_5$ | $< 5.9 \times 10^{-21}$ | 47.658° N, 122.3° W | Electron | $\hat{\beta}^e_N$ |
| [16] | AlNiCo-SmCo$_5$ | $< 8 \times 10^{-22}$ | 47.658° N, 122.3° W | Electron | $\beta^e_E$ |
| [39] | $^{199}$Hg-$^{133}$Cs | $< 4.3 \times 10^{-20}$ | 42.37° N, 72.53° W | Proton | $\hat{\beta}^p_N$ |
| [17] | $^{199}$Hg-$^{133}$Cs | $< 3 \times 10^{-20}$ | 42.37° N, 72.53° W | Proton | $\beta^p_E$ |
| [21] | $^{85}$Rb-$^{87}$Rb | $<1.3 \times 10^{-18}$ | 37.66° N, 122.05° W | Proton | $\beta^p_z$ |
| [17] | $^{199}$Hg-$^{133}$Cs | $< 2.9 \times 10^{-21}$ | 42.37° N, 72.53° W | Neutron | $\beta^n_E$ |
| [22] | $^{129}$Xe-$^{131}$Xe | $<2.3 \times 10^{-22}$ | 31.82° N, 117.23° E | Neutron | $\hat{\beta}^n_z$ |

**TABLE I.** Best experimental bounds on orientation-dependent energy shifts $\beta$. The subscript denotes the spin orientation (see text). All bounds are given at the 95% C.L.

All of these results are compiled in Table 1. For the p-N results we have included the results of both Ref. [39] and Ref. [21] as the former depends upon nuclear structure calculations whereas the latter does not. A review of the relevant nuclear spin content is presented in Ref. [41]. A hat over beta indicates that a correction has been applied to the data to account for the gyro-compass effect due to Earth's rotation.

To establish bounds on the coupling coefficients we require that the associated energy shift of the total potential ($V_T$) be less than the energy bound established on the spin-coupling energy (β) in the spin-sensitive direction for the various experiments. We assume there is no cancellation of the effect by other exotic potentials. The resulting bounds on the potentials are shown in Fig. 1 for $\lambda > 100$ m. This work's constraints given in graphical form are at the 95% confidence level ($2\sigma$). This work improves bounds on the e-N coupling constants $f_{4+5}$ and $f_{12+13}$ by as much as 18 and 8 orders of magnitude, respectively, for the range $\sim 10^3$ m $< \lambda < 10^{10}$ m. In the same range, we improve the bounds on $f_{4+5}$ for n-N and p-N couplings by as much as 9 and 8 orders of magnitude, respectively. The n-N coupling $f_{12+13}$ is improved by as much as 5 orders of magnitude in that range. We are unaware of any previous long-range bounds on the p-N $f_{12+13}$ coupling. Other measurements of individual scalar, vector, and axial coupling constants can be interpreted [42, 43, 44] to constrain these couplings as well. Given the wide range of exotic interactions that have been proposed, direct bounds on each of the allowed potentials remain valuable.

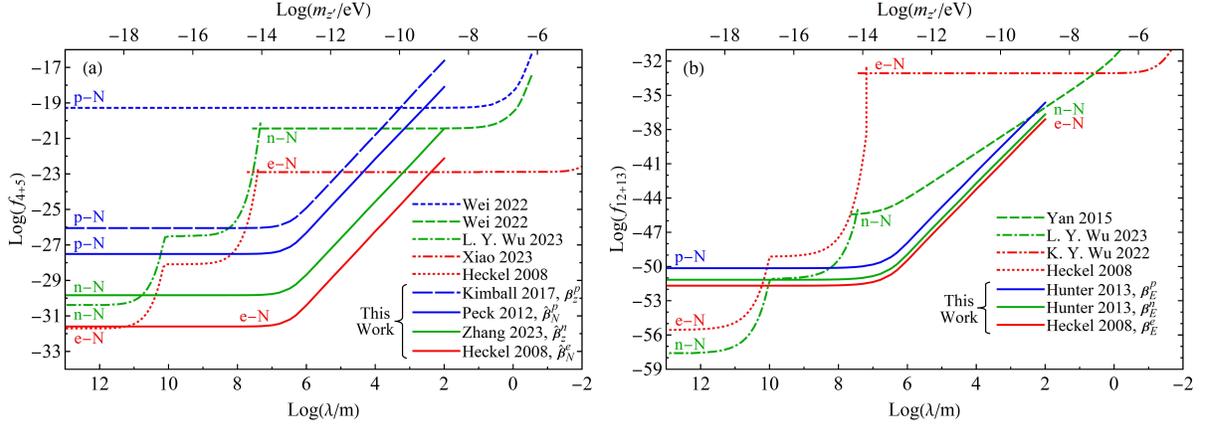

**FIG. 1.** Upper limits on the $f_{4+5}$ and $f_{12+13}$ coupling constants as a function of interaction range. Each curve is identified using the convention that the first fermion from each labeled pair represents the spin-polarized particle and the second the unpolarized particle. Previous e-N (red) constraints of K. Y. Wu [13], Xiao [14], and Heckel [16], n-N (green) constraints of Yan [19], L. Y. Wu [23], and Wei [24], and p-N (blue) of Wei [24] are shown with dashed lines. In (a), this work's e-N (red), n-N (green), and p-N (blue) constraints using bounds extracted from $\hat{\beta}^e{}_N$ [16], $\hat{\beta}^n{}_z$ [22], $\beta^p{}_z$ [21], and $\hat{\beta}^p{}_N$ [39] are shown with solid lines save for bounds extracted from [21] shown as long dashes. In (b), this work's e-N (red), n-N (green), and p-N (blue) constraints using bounds extracted respectively from $\beta^e{}_E$ [16], $\beta^n{}_E$ [17], and $\beta^p{}_E$ [17] are presented as solid lines. The region above these constraints is excluded at 95% ($2\sigma$) confidence, save for the work of Ref. [13] and [14] which excludes at the 68% ($1\sigma$) confidence level. The bounds for all $m_{z'} < 10^{-20}$ eV are the same as those displayed at $m_{z'} = 10^{-20}$ eV.

## IV. DISCUSSION

The slopes of our exclusion lines in Fig. 1 between $10^2$ m $< \lambda < 10^6$ m can be roughly understood from dimensional analysis. In this range as $\lambda$ increases the number of particles sampled increases as $\lambda^3$ (neglecting geometric and density changes). The explicit radial dependence of the newly included geoparticles drops off approximately as $1/r$ for $V_{12+13}$ and $1/r^2$ for $V_{4+5}$. However, their relative velocities increase as $r$, so the sensitivity increases proportional to $\lambda^3$ and $\lambda^2$ for $V_{12+13}$ and $V_{4+5}$ respectively, as is approximately observed in Fig. 1. A similar argument applied to the slope of the exclusion line of Ref. [19], where the velocity does not increase with distance, yields a sensitivity increase proportional to $\lambda^2$ for $V_{12+13}$ as is illustrated in Fig. 1.

The dimensionless coupling constants $f_{4+5}$ and $f_{12+13}$ can be given in terms of scalar ($g_S$), vector ($g_V$), and axial ($g_A$) coupling constants for the case of single massive spin-0 or spin-1 boson exchange following Ref. [10, 45]. We began our analysis by assuming the polarized particle in $f_{4+5}$ and $f_{12+13}$ only interacts with nucleons. Then for the case of spin-1 boson exchange $f_{12+13} = 4g_A^i g_V^N$, where the $i$ superscript specifies the polarized particle and the other superscripts specify the unpolarized fermion species. The $f_{12+13}$ coupling constant is zero for single spin-0 boson exchange. The same convention for $f_{4+5}$ yields $f_{4+5} = g_S^i g_S^N$ for spin-0 boson exchange and $f_{4+5} = -(g_A^i g_A^N + 3g_V^i g_V^N)/2$ for spin-1 boson exchange.

Constraints for other choices of couplings can be readily obtained by scaling our limits. For instance, one can estimate the constraints on $g_A^i g_V^e$ or $g_A^i g_V^p$ rather than $g_A^i g_V^N$ by multiplying the relevant nucleon bound by ~2.06 (the inverse of the average proton/nucleon ratio of Earth). Similarly, the constraints on $g_A^i g_V^n$ can be determined by multiplying the relevant combined nucleon bounds by ~1.94 = 1/(1-0.486). This conversion factor is not exact because the unpolarized particle density of Earth, $\rho(r')$, depends on $r'$ and cannot be simply factored out of the volumetric integrals (Eq. 3 and Eq. 4). As such the conversion factor is not a constant but depends on the details of the integration. We have carried out the integrations with our best estimates of the number of each species of particle present (e, n and p, based on the elemental composition of Earth) as a function of their distance from the center of Earth. The results of this more accurate integration method agree with the above approximation to better than 2%.

## V. CONCLUSIONS

We have combined the results from various experiments with an Earth model to obtain bounds on the long-range spin-velocity couplings $f_{4+5}$ and $f_{12+13}$ that couple the spins of electrons, neutrons, and protons to the velocities of moving electrons, neutrons, protons and nucleons.

Improved measurements of the energies associated with various fermion-spin orientations relative to Earth will further improve bounds on these long-range spin-velocity couplings, as well as both velocity-independent [17] and velocity-dependent [18] spin-spin couplings. A new generation of the Amherst investigation [17] uses free-precession magnetometers and hopes to achieve an order-of-magnitude improvement in sensitivity. In the future, using Earth as a source of moving particles should continue to provide a valuable means of constraining $f_{4+5}$ and $f_{12+13}$ at ranges greater than 100 m.

## ACKNOWLEDGMENTS

This work was supported by National Science Foundation Grant No. PHY-2110523. The authors thank D. Budker, L. Cong, A. Glassford, B. Herz, D. F. Jackson Kimball, W. Loinaz, S.K. Peck, S. Salim, and N.T. Akinci for useful conversations.